\documentclass{article}

\newcommand{\bls}[1]{\renewcommand{\baselinestretch}{#1}}
\def\noi{\noindent}

\makeatletter
\renewcommand{\section}{\@startsection{section}{1}{0pt}%
        {-3.5ex plus -1ex minus -.2ex}{2.3ex plus .2ex}%
        {\large\bf\protect\raggedright}}

\renewcommand{\subsection}{\@startsection{subsection}{2}{0pt}%
        {-3ex plus -1ex minus -.2ex}{1.4ex plus .2ex}%
        {\normalsize\bf\protect\raggedright}}

\renewcommand{\@oddhead}{\raisebox{0pt}[\headheight][0pt]{%
   \vbox{\hbox to\textwidth{\rightmark \hfil \rm \thepage \strut}\hrule}}}
\renewcommand{\@evenhead}{\raisebox{0pt}[\headheight][0pt]{%
   \vbox{\hbox to\textwidth{\thepage \hfil \leftmark \strut}\hrule}}}

\makeatother

\newcommand{\Title}[1]{\noi {\Large #1} \\}
\newcommand{\Author}[2]{\noi{\large\bf #1}\\[2ex]\noindent{\it #2}\\}
\newcommand{\Abstract}[1]{\vskip 2mm \begin{center}
        \parbox{16.4cm}{\small\noi #1} \end{center}\medskip}

\newcommand{\PACS}[1]{\begin{center}{\small PACS: #1}\end{center}}
\newcommand{\foom}[1]{\protect\footnotemark[#1]}

\newcommand{\email}[2]{\footnotetext[#1]{e-mail: #2}}

\newcommand{\Ref}[1]{Ref.\,\cite{#1}}

\def\nqq{\hspace*{-2em}}

\def\nhh{\hspace*{-0.3em}}
\def\cm{\hspace*{1cm}}
\def\inch{\hspace*{1in}}
\newcommand{\Theorem}[2]{\medskip\noi {\bf #1. \ }{\it #2}\medskip}
\newcommand{\Picture}[3]{
	\begin{figure} 	\centering \unitlength=1mm
	\begin{picture}(86,#1)
		\put(0,0){\line(0,1){#1}}            %frame
		\put(0,0){\line(1,0){86}}
		\put(86,0){\line(0,1){#1}}
		\put(0,#1){\line(1,0){86}}
	\put(6,0){#2}                       \end{picture}
        \caption{\protect\small #3}  \smallskip \hrule \end{figure} }

\def\eq{Eq.\,}
\def\eqs{Eqs.\,}
\def\beq{\begin{equation}}
\def\eeq{\end{equation}}
\def\bear{\begin{eqnarray}}
\def\al{&\nhh}
\def\lal{&&\nqq {}}               % left alignment
\def\bearr{\begin{eqnarray} \lal}
\def\ear{\end{eqnarray}}
\def\earn{\nonumber \end{eqnarray}}

\def\tst{\textstyle}

\def\nn{\nonumber\\ {}}

\def\nnn{\nonumber\\ \lal }

\def\yy{\\[3pt] {}}
\def\yyy{\\[5pt] \lal }
\def\eql{\al =\al}

\def\eqdef{\stackrel{\rm def}=}
\def\e{{\,\rm e}}
\def\d{\partial}
\def\sign{\mathop{\rm sign}\nolimits}
\def\diag{\mathop{\rm diag}\nolimits}

\def\const{{\rm const}}

\def\half{{\tst\frac{1}{2}}}
\def\then{\ \Rightarrow\ }

\def\bh{black hole}
\def\bhs{black holes}
\def\sph{spherically symmetric}
\def\mn{_{\mu\nu}}
\def\mN{_{\mu}^{\nu}}
\def\MN{^{\mu\nu}}
\def\nM{^{\mu}_{\nu}}

\def\H{{\cal H}}
\def\eff{_{\rm eff}}

\def\RN{Reissner-Nordstr\"om}
\def\fe{f_{\rm e}}
\def\fm{f_{\rm m}}
\def\qe{q_{\rm e}}
\def\qm{q_{\rm m}}

\bls{0.975}

\begin{document}
\thispagestyle{empty}
\twocolumn[

\rightline{\bf gr-qc/0006014}
\medskip

\Title
  {\bf Regular magnetic black holes and monopoles \yy
       from nonlinear electrodynamics}

\Author{K.A. Bronnikov\foom 1}
{Centre for Gravitation and Fundam. Metrology, VNIIMS,
        3-1 M. Ulyanovoy St., Moscow 117313, Russia;\\
Institute of Gravitation and Cosmology, PFUR,
        6 Miklukho-Maklaya St., Moscow 117198, Russia}

\Abstract
{It is shown that general relativity coupled to nonlinear electrodynamics
(NED) with the Lagrangian $L(F)$, $F = F\mn F\MN$ having a correct weak
field limit, leads to nontrivial \sph\ solutions with a globally regular
metric if and only if the electric charge is zero and $L(F)$ tends to a
finite limit as $F \to \infty$.  Properties and examples of such solutions,
which include magnetic black holes and solitonlike objects (monopoles), are
discussed. Magnetic solutions are compared with their electric
counterparts. A duality between solutions of different theories specified in
two alternative formulations of NED (called the $FP$ duality) is used as a
tool for this comparison.  \PACS{04.20.Jb, 04.20.Dw, 04.70.Bw} }

] %%%%%%%%%%%%%%%%%%%%%%%%%%%
\email 1 {kb@rgs.mccme.ru}

\section{Introduction}

    General relativity, despite its nonlinearity, is apparently lacking an
    effective self-restriction mechanism, and the existence of singularities
    seems to be its inevitable, though undesired, feature. Reasonable,
    regular solutions for macroscopic bodies such as stars are obtained with
    matter whose pressure opposes gravity, whereas microobjects, extreme
    states of matter and/or strong gravitational fields probably need a
    purely field description.

    The choice of a field source able to do the job is a separate task,
    and, in particular, for \sph\ configurations there are quite a number of
    nonexistence theorems \cite{noex}. Non-Abelian gauge
    fields yield regular \bh\ solutions \cite{YM}, but they are known only
    numerically. The regular \bh\ solution of \Ref{dym1} with a de Sitter
    core is expressed in terms of pressure and density rather than fields.
    An especially attractive class of field theories for seeking regular
    models is nonlinear electrodynamics (NED) with gauge-invariant
    Lagrangians $L(F)$, $F=F\mn F\MN$, since its energy-momentum tensor
    (EMT) $T\mN$ has the symmetry $T_0^0 = T_1^1$ and is thus insensitive to
    boosts in the radial direction, which is a property of vacuum
    \cite{dym1,dym2}.  Such theories, in particular, the Born-Infeld
    NED, recently gained much attention as limiting cases of certain
    models of string theory (see \cite{seywitt} for reviews). It has been
    shown, however \cite{b76,b79}, that in NED with any $L(F)$ such that
    $L\sim F$ at small $F$ (the Maxwell weak-field limit), static, \sph\
    general-relativistic configurations with a nonzero electric charge
    cannot have a regular center.  As will be shown below, the same is true
    for dyonic configurations, where both electric and magnetic charges are
    present; regular solutions with wormhole topology also cannot exist for
    this system.

    The prohibition does not concern purely magnetic solutions, and,
    quite surprisingly, there is a whole class of regular solutions with a
    nonzero magnetic charge. Regularity is here understood in a physical
    sense: although $F$ is infinite at the center, the EMT is finite and the
    metric is regular (at least $C^2$) while the forces applied to test
    particles are finite everywhere and vanish at the center. The main aim
    of this paper is to present and to discuss these solutions. We will also
    compare them with their electric analogs, in particular, with the
    solutions found by Ay\'on-Beato and Garc\'\i a [8--10]. For this
    comparison we use a duality between \sph\ solutions of different
    NED specified in two alternative ($F$ and $P$) frameworks:  the
    original, Lagrangian framework and another one, obtained from it by a
    Legendre transformation \cite{sala}.

    Throughout the paper all relevant functions are assumed to be
    sufficiently smooth, unless otherwise indicated.

\section{Nonexistence theorems}

    Let us begin with a proof of two simple nonexistence theorems
    extending the results of \cite{b76,b79}.

    Consider NED in general relativity, with the action
\beq                                                            \label{e1}
    S = \frac{1}{16\pi}
      \int d^4 x\sqrt{-g}\ [R - L(F)], \ \  F \eqdef F\mn F\MN,
\eeq
    where $R$ is the scalar curvature, $F\mn = \d_\mu A_\nu- \d_\nu A_\mu$
    is the electromagnetic field, and $L$ is an arbitrary function leading
    to the Maxwell theory at small $F$: $L(F)\approx F$ as $F\to 0$. The
    tensor $F\mn$ obeys the dynamic equations and the Bianchi identities,
\beq
    \nabla_\mu(L_F F\MN)=0, \cm  \nabla_\mu {}^*F\MN =0,       \label{dynF}
\eeq
    where an asterisk denotes the Hodge dual and $L_F = dL/dF$.

    A static, \sph\ metric can be written in the general form
\beq                                                            \label{e2}
     ds^2 = \e^{2\gamma(u)}dt^2 - \e^{2\alpha(u)}du^2 - r^2(u)\,d\Omega^2
\eeq
    where $d\Omega^2 = d\theta^2 + \sin^2\theta\, d\phi^2$ and $u$ is a
    radial coordinate. A regular center, by definition, takes place where
    $r=0$, if all algebraic curvature invariants are finite there and, in
    addition, there is a correct limiting value of the circumference to
    radius ratio (local flatness at the center, the absence of a conical
    singularity). If one chooses the curvature coordinates, $u=r$, the local
    flatness condition reads $\e^{\alpha (0)}=1$.

    The tensor $F\mn$ compatible with spherical symmetry can involve
    only a radial electric field $F_{01}=-F_{10}$ and a radial magnetic
    field $F_{23}=-F_{32}$. \eqs(\ref{dynF}) give
\beq
    r^2 \e^{\alpha+\gamma}L_F F^{01} = \qe, \cm
                                  F_{23} = \qm \sin\theta       \label{qq}
\eeq
    where $\qe$ and $\qm$ are the electric and magnetic
    charges, respectively. As follows from (\ref{qq}),
\bear                                                             \label{fe}
    \fe \eqdef 2F_{01}F^{10} \eql  2\qe^2\,L_F^{-2} r^{-4} \geq 0, \\
    \fm \eqdef 2F_{23}F^{23} \eql  2\qm^2\,r^{-4} \geq 0,     \label{fm}
\ear
    and the Einstein equations may be written in the form
\bearr                                                            \label{EE}
 -G\mN = T\mN = -2 L_F F_{\mu\alpha}F^{\nu\alpha}
                                                +\half \delta\mN L
    \\     \lal                                                  \label{EMT}
  = \half \diag
    ( L {+} 2\fe L_F,\, L {+} 2\fe L_f,\,  L{-}2\fm  L_F,\,  L{-}2\fm L_F).
    \nnn
\ear

\Theorem{Theorem 1}
    {The field system (\ref{e1}), with $ L(F)$ having a Maxwell asymptotic
    ($ L\to 0,\  L_F\to 1$ as $F\to 0$),
    does not admit a static, \sph\ solution with a regular center and a
    nonzero electric charge.}

\noi
    {\bf Proof.}
    Since the Ricci tensor for the metric (\ref{e2}) is diagonal,
    the invariant $R\mn R\MN \equiv R\mN R\nM$ is a sum of
    squares, hence each component $R_\mu^\mu$ (no summing) is finite
    at a regular space-time point. Then each component of the
    EMT $T\mN$ is finite as well, hence, as follows from (\ref{EE}),
\beq
        (\fe + \fm) | L_F| < \infty.                       \label{fin}
\eeq

    Suppose first that $\qm=0,\ \qe\ne 0$ and thus $\fm=0$ and $F = -\fe$.
    Therefore, by (\ref{fe}) and (\ref{fin}), at a regular center $F L_F$ is
    finite whereas $F L_F^2 \to \infty$. Combined, these conditions lead to
    $F\to 0$ and $L_F\to\infty$, that is, a strongly non-Maxwell behavior at
    small $F$. Thus for purely electric fields the theorem is valid.

    Suppose now $\qe\ne 0$ and $\qm\ne 0$. Then (\ref{fin}) should hold for
    $\fe$ and $\fm$ taken separately. As stated previously, this condition
    applied to $\fe$ combined with (\ref{fe}) leads to $ L_F\to \infty$. But
    $\fm$ also tends to infinity as $r\to 0$, so even stronger $\fm L_F\to
    \infty$, violating (\ref{fin}). The theorem is proved.

\medskip
    A regular center is, however, not a necessary feature of a regular
    static, \sph\ space-time: there might be no center at all. Let us find
    out whether or not our system (\ref{e1}) can behave like this.

    If a center is lacking, the metric function $r=(-g_{22})^{1/2}$ is
    restricted below by some minimum value $r^* > 0$.  There are then two
    opportunities:

\begin{description}
\item[(i)]
    $r(u)$ has one or several minima (as it happens, e.g., in a wormhole);
    $r$ itself then fails to be an admissible coordinate since it takes
    equal values at different spheres;
\item[(ii)]
    $r(u)$ monotonically approaches $r_{\min}$ as $u$ tends to a certain
    limiting value $u^*$; then the space-time can be nonsingular if it
    ends with a horn.
\end{description}

    By definition, {\it a {\bf horn} in a static, \sph\ space-time with the
    metric (\ref{e1}) is a region where, as $u$ tends to some value $u^*$,
    $r(u)\ne \const$ and $g_{tt}=\e^{2\gamma(u)}$ have finite limits while
    the integral $\ell = \int \e^\alpha du$ diverges.\/} In other words, a
    horn is an infinitely long (3-dimensional) ``tube'' of finite radius,
    with the clock rate remaining finite everywhere. It has an infinite
    spatial volume, and geodesics are infinitely continued along it as if in
    a wormhole throat of unlimited length. The condition $r\ne \const$
    discards purely cylindrical space-times, sometimes called flux tubes,
    which have no asymptotics. The definition of a horn follows the papers
    by Banks et al. \cite{banks}, where ``horned particles'' were discussed
    as possible remnants of black hole evaporation.

    The following theorem shows that these opportunities can never be
    realized if the EMT has the ``vacuum'' property $T^0_0 = T^1_1$, and, in
    particular, for our system (\ref{e1}).

\Theorem{Theorem 2}
    {Let the metric (\ref{e2}) obey the Einstein equations with an EMT
    satisfying the condition $T^0_0 = T^1_1$. Then (i) the function $r(u)$
    cannot have a regular minimum and (ii) the space-time cannot contain a
    horn.}

\noi
    {\bf Proof.}
    Let us choose the $u$ coordinate by fixing the condition
    $\alpha=\ln r - \gamma$. The ${0\choose 0}-{1\choose 1}$
    Einstein equation takes the form $d^2(\ln r)/du^2=0$. Therefore
\beq
    r(u) = \e^{au+b}, \cm a,\,b=\const,                    \label{ru}
\eeq
    so that $r$ is either a constant (if $a=0$) or a strictly monotonic
    function. This proves item (i).

    Suppose now that there is a horn. Then, by the above
    definition, $a\ne 0$, and a finite limit of $r$ in \eq (\ref{ru}) as
    $u\to u^*$ means that $u^*$ is also finite. On the other hand, since
    $\int \e^\alpha du$ diverges as $u\to u^*$, it follows that $\alpha\to
    \infty$, which, by our coordinate condition, can only occur if
    $\gamma\to -\infty$, contrary to what was assumed. This completes the
    proof.

\medskip
    Some remarks are now in order. First, the absence of wormhole
    solutions can also be proved from the known fact that a static wormhole
    throat always implies a violation of the null energy condition
    \cite{hovis}.  This condition in our case reads $T_0^0 - T^1_1 \geq 0$
    and is (marginally) observed by the identical zero at the left-hand
    side. Our proof is, however, more direct and explicit.

    Second, the opportunity $\gamma \to -\infty$, mentioned in the proof of
    item (ii) of Theorem 2, generically corresponds to an event horizon.
    [It can happen in principle that $\e^\gamma\to 0$ at finite $r$ does not
    imply a horizon. Our system (\ref{e1}) with smooth $L(F)$ does not
    admit such cases, while horizons do exist, as can be seen from the
    solution below.]

    Third, the above theorems do not contain an asymptotic flatness
    requirement, the proofs being of local nature. Therefore both theorems
    are readily extended to general relativity with a cosmological constant
    (the condition $T_0^0=T_1^1$ holds in its presence as well),
    where the spatial asymptotic can be de Sitter or anti-de Sitter.

\section{Regular magnetic solutions}

    Both theorems were proved without entirely solving the Einstein
    equations. For our system (\ref{e1}), however, an exact solution can
    be obtained by quadratures in the general \sph\ case \cite{b76}. Indeed,
    the Maxwell-like equations are already integrated. Let us choose the
    curvature coordinates $u=r$, which is now safe since we know that $r$
    has no extrema. Due to $T^0_0 = T^1_1$, the corresponding Einstein
    equation gives
    		$d(\alpha+\gamma)/dr=0 \then \alpha+\gamma =0$
    for a proper choice of the time unit. It remains to write the
    well-known relation for $\alpha(r)$ in terms of the energy density
    $T_0^0$, which follows from the ${0\choose 0}$ Einstein equation:
\bear                                                            \label{M}
    \e^{2\gamma} = \e^{-2\alpha} \eql A(r) = 1 - \frac{2M(r)}{r}, \nn
     M(r) \eql
        \frac{1}{2} \int T^0_0(r)\, r^2\,dr.
\ear
    Possible horizons occur at zeros of $A(r)$.

    The only nontrivial case not covered by the theorems is $\qe=0$,
    $\qm\ne 0$, when there is still a hope to obtain a regular center.
    In this case the metric has the form (\ref{e2}) with (\ref{M}) where
\beq
                  M(r)=\frac{1}{4}\int L(F) r^2 dr          \label{ML}
\eeq
    and $F = 2\qm^2/r^4$. It is easily seen that a space-time with a regular
    center is indeed obtained for any $L(F)$ such that $ L \to  L_\infty <
    \infty$ as $F\to \infty$ when one integrates in \eq (\ref{ML}) from 0 to
    $r$.  Integration from 0 to $\infty$ then gives a unique mass
    $m=M(\infty)$ providing a regular center for given $\qm$; hence the
    entire mass is of electromagnetic origin. A finite value of $M(\infty)$
    is guaranteed by our assumption of a Maxwell asymptotic of
    $L(F)$ at small $F$: as $r\to \infty$, $L\approx 2\qm^2/r^4$, and the
    integral (\ref{ML}) converges.

    The EMT in (\ref{EE}) near $r=0$ takes the form $T\mN = \half\delta\mN
    L_\infty (1+o(1))$, and the metric is approximately de Sitter ($A(r) = 1
    - \Lambda r^2/3 + o(r^2)$), with the cosmological constant $\Lambda=
    L_\infty/2$. The Riemann tensor at $r=0$ coincides with that of de
    Sitter space, so one need not explicitly calculate the curvature
    invariants to prove that the space-time is regular. The metric is at
    least $C^2$ smooth at $r=0$ but, depending on $L(F)$, may be even
    $C^\infty$ smooth, as will be seen in the example (\ref{g2m}).

    Suppose that $L(F)$ and the mass have been chosen as described.
    The space-time is then globally regular and can include horizons
    corresponding to zeros of $A(r)$, whose number and character determine
    the global structure [note that $A(0)=A(\infty)=1$]. In the absence of
    zeros, there is a regular Dirac-type magnetic monopole solution. The
    occurence of two simple zeros leads to the conventional
    \RN\ \bh\ structure, with the singularity replaced by a regular
    center. The intermediate case of one double zero gives the extremal \RN\
    structure. Models with more numerous horizons can be constructed as well.

    Due to the above theorems, our {\it magnetic \bhs\ and monopoles
    are the only types of static, \sph\ solutions to (\ref{e1}) with a
    globally regular metric}.

    The only trouble is an infinite magnetic induction $B$ ($B^2 =
    \fm/2 =F/2$) at $r = 0$, whereas the magnetic field intensity $H$,
    obtained as a generalized momentum from the Lagrangian, is well-behaved
    everywhere including the center: $H^2 = \fm L_F^2/2\to 0$ as $r\to 0$.
    To judge whether or not the center is regular from a physical viewpoint,
    one should estimate the force applied to a charged test particle
    moving in the field under consideration. This test charge may be
    electric or magnetic since both are admitted by our assumptions. In
    a consistent approach, the equations of motion for a test particle (as
    well as those for an extended body) in nonlinear field theory should
    follow from the field equations and may be deduced along the lines of
    Refs.\,\cite{force,ryb}. Namely, the 4-force vector is found
    as the total momentum flow $\int T\mn n^\mu dS$ through a closed surface
    surrounding the particle, where $n^\mu$ is the unit normal to such a
    surface and $T\mn$ is the total EMT of the summed electromagnetic field
    of the background static configuration and the test particle itself. An
    estimate in a proper approximation, taking into account the weakness of
    the particle's field, shows that this force is finite everywhere and
    vanishes at $r=0$. Therefore, despite the divergent $B$, our
    magnetic solutions may be called globally regular.

\section{$FP$ duality and electric solutions}

    Let us consider for comparison the electric analogs of our magnetic
    solutions. This is of particular interest since recently Ayon-Beato and
    Garc\'{\i}a [8--10] suggested some examples of such
    solutions, describing configurations with $\qe \ne 0$, $\qm =0$ and a
    regular center. The properties of these solutions evidently contradict
    Theorem 1, but they only seem to circumvent it since, as we shall see,
    any model like those of [8--10] needs different Lagrangians in different
    ranges of the radial coordinate and therefore fails to be a solution for
    a particular Lagrangian $L(F)$.

    The solutions of [8--10] were found using an alternative
    form of NED (to be called the $P$ framework),
    obtained from the original one (the $F$ framework)
    by a Legendre transformation: one introduces the tensor $P\mn =  L_F
    F\mn$ with its invariant $P = P\mn P\MN$ and considers the
    Hamiltonian-like quantity $\H= 2F L_F - L$ as a function of $P$; the
    theory is then reformulated in terms of $P$ and is specified by
    $\H(P)$ \cite{sala}. One then has:
\beq
    L = 2P\H_P-\H, \quad   L_F \H_P =1, \quad F = P\H_P^2      \label{LH}
\eeq
    with $\H_P=d\H/dP$. \eqs (\ref{dynF}) and the EMT (\ref{EE})
    are rewritten in the form
\bearr
    \nabla_\mu P\MN=0, \cm  \nabla_\mu (\H_P\,{}^*F\MN) = 0,  \label{dynP}
    \yyy
T\mN = -\half \diag                                           \label{EMT-P}
   ( \H {-} 2p_{\rm m} \H_P,\, \H {-} 2p_{\rm m} \H_P,\,\nnn
    \inch  \cm
    		\H {+} 2p_{\rm e} \H_P,\, \H {+} 2p_{\rm e} \H_P)
\ear
    where, by (\ref{dynP}), in the \sph\ case,
\bear                                                         \label{pe}
    p_{\rm e} \eql 2 P_{01}P^{10} = 2\qe^2/r^4, \\         \label{pm}
    p_{\rm m} \eql 2 P_{23}P^{23} = 2\qm^2 \H_P^2/r^4.
\ear

    Comparing \eqs (\ref{dynF}), (\ref{fe})--(\ref{EMT}) with
    (\ref{dynP})--(\ref{pm}), one sees that they coincide up to the
    substitutions
\beq
    \{ g\mn,\, F\mn,\, F,\, L(F)\} \ \longleftrightarrow \
    \{ g\mn,\, {}^* P\mn,\, -P,\, -\H(P)\}.
\eeq
    In other words, there is a duality between \sph\ solutions written in
    the $F$  and $P$ frameworks: any solution for a given Lagrangian
    $L(F)$, characterized by a certain metric function $A(r)$ and the field
    components $F_{01}$ and $F_{23}$, has a counterpart with the same
    $A(r)$ but with $F$ substituted by $-P$, $L$ by $-\H$, $F_{01}$ by
    $P_{23}$ and $F_{23}$ by $P_{01}$, and conversely. The functional
    dependence $-\H(-P)$ in the dual solution is the same as $L(F)$ in the
    original solution, but the choice of the function $L(F)$ itself is not
    restricted.

    (It should be stressed that this duality, to be called $FP$ {\it
    duality\/}, connects solutions of {\it different\/} theories:  given
    $L(F)$, the functional dependence $\H(P)=2FL_F -L$ is in general quite
    different from $L(F)$, an evident exception being the Maxwell theory,
    where $L=F=\H=P$ and the present duality turns into the conventional
    electric-magnetic duality.  So the $FP$ duality has nothing to do with
    the electric-magnetic one studied in Refs.\,\cite{sala,gibrash}, where
    the field equations of a specific theory were required to be duality
    invariant, and this condition selected a narrow class of Lagrangians.)

    In particular, any regular magnetic solution obtained for given $L(F)$
    has a purely electric counterpart with a similar (up to the sign)
    dependence $\H(P)$. The metric has the form (\ref{M}) with
\beq
                  M(r)=-\frac{1}{4}\int \H(P) r^2 dr.            \label{MH}
\eeq
    Given $\H(P)$, one should substitute $P=-2\qe^2/r^4$.
    A regular center exists if and only if $\H$ has a finite limit as
    $P\to -\infty$, and a mass that provides regularity for given $\qe$ is
    found by integration in (\ref{MH}) from 0 to $\infty$. This is how the
    regular solutions of [8--10] were obtained with the following choices
    of $\H(P)$:
\bear
    \cite{gar1}:
        \H(P) \eql P \frac{1-3\Pi}{(1+\Pi)^3}         \label{g1}
                        + \frac{6}{q^2 s}
                \biggl(\frac{\Pi}{1+\Pi}\biggr)^{5/2},       \\  \label{g2}
    \cite{gar2}:
            \H(P) \eql  P \big/ \cosh^2 (s\sqrt{\Pi}),   \\  \label{g3}
    \cite{gar3}:
            \H(P) \eql  P \frac{\exp(-s\sqrt{\Pi})}{(1+\Pi)^{5/2}}
                         \biggl(1 + \frac{3}{s}\sqrt{\Pi} + \Pi\biggr),
\ear
    where $\Pi = \sqrt{-q^2 P/2}$ and $s = |q|/(2m)$, $q=\qe$ and $m$ being
    free parameters identified with the charge and mass of the
    configuration, respectively. The functions (\ref{g1})--(\ref{g3}) behave
    like $P$ at small $P$, tend to finite limits as $P\to -\infty$
    and thus lead to regular metrics.
\Picture{73}
{
\unitlength 0.45mm
\linethickness{0.4pt}
\begin{picture}(158.00,152.00)
\put(6.00,92.00){\vector(0,1){60.00}}
\put(1.00,97.00){\vector(1,0){70.00}}
\put(6.00,15.00){\vector(0,1){55.00}}
\put(1.00,20.00){\vector(1,0){70.00}}
\put(98.00,20.00){\vector(0,1){120.00}}
\put(50.00,135.00){\makebox(0,0)[cc]{$\H(P)$}}
\put(50.00,63.00){\makebox(0,0)[cc]{$F(P)$}}
\put(129.00,12.00){\makebox(0,0)[cc]{$L(F)$}}
\put(73.00,91.00){\makebox(0,0)[cc]{$-P$}}
\put(73.00,14.00){\makebox(0,0)[cc]{$-P$}}
\put(16.00,148.00){\makebox(0,0)[cc]{$-\H$}}
\put(17.00,68.00){\makebox(0,0)[cc]{$-F$}}
\put(108.00,140.00){\makebox(0,0)[cc]{$-L$}}
\put(93.00,117.00){\vector(1,0){65.00}}
\bezier{412}(6.00,20.00)(16.67,78.00)(29.00,36.00)
\bezier{212}(29.00,36.00)(35.00,7.00)(41.00,30.00)
\bezier{212}(41.00,29.00)(46.67,53.33)(58.00,28.00)
\bezier{320}(6.00,97.00)(29.67,140.00)(47.00,115.00)
\bezier{124}(47.00,115.00)(57.00,101.00)(71.00,99.00)
\bezier{208}(98.00,117.00)(138.67,122.00)(149.00,127.00)
\bezier{692}(149.00,127.00)(61.00,80.00)(125.00,45.00)
\bezier{348}(128.00,43.00)(97.00,74.00)(98.00,117.00)
\put(37.00,93.00){\makebox(0,0)[cc]{$P_2$}}
\put(36.00,14.00){\makebox(0,0)[cc]{$P_2$}}
\put(21.00,14.00){\makebox(0,0)[cc]{$P_1$}}
\put(51.00,14.00){\makebox(0,0)[cc]{$P_3$}}
\put(144.00,130.00){\makebox(0,0)[cc]{$P_1$}}
\put(93.00,74.00){\makebox(0,0)[cc]{$P_2$}}
\put(127.00,38.00){\makebox(0,0)[cc]{$P_3$}}
\put(157.00,111.00){\makebox(0,0)[cc]{$-F$}}
\bezier{56}(58.00,28.00)(60.67,23.00)(69.00,22.00)
\bezier{16}(125.00,45.00)(125.67,44.00)(128.00,43.00)
\linethickness{0.15pt}
\put(34.00,124.00){\line(0,-1){27.00}}
\put(34.00,97.00){\line(0,1){0.00}}
\put(19.00,54.00){\line(0,-1){34.00}}
\put(48.00,41.00){\line(0,-1){21.00}}
\put(93.00,112.00){\makebox(0,0)[cc]{$O$}}
\end{picture}
}
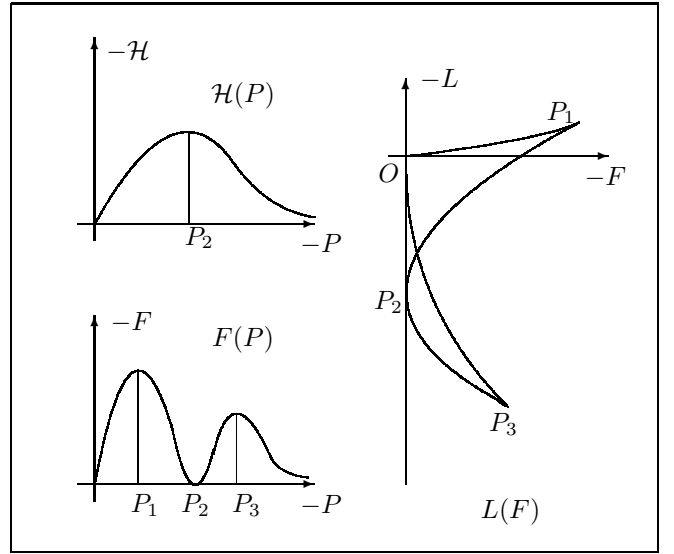
{An example of qualitative behavior of $\H(P)$,\ $F(P)$, and $ L(F)$
          in an electric solution}

    Yet the $P$ framework is secondary: the Lagrangian dynamics is
    specified in the $F$ framework, and, since the $F\mapsto P$ transition
    is a mere substitution in the field equations, the two frameworks are
    only equivalent where the function $P(F)$ is monotonic. Recalling the
    proof of Theorem 1 for $\qe\ne 0$, $\qm=0$, one sees, however, that for
    any regular solution with a \RN\ asymptotic the function $F(P) = -\fe
    \leq 0$ vanishes at both $r=0$ and $r\to \infty$. So it inevitably has
    at least one minimum at some $P=P^* < 0$. It can be shown directly that
    at an extremum of $F(P)$ where $F = F^* < 0$ the derivative $ L_F$ has
    the same finite limit as $P\to P^*+0$ and $P\to P^*-0$, while $ L_{FF}$
    tends to infinities of opposite signs.  Therefore the function $ L(F)$
    suffers branching, and its graph forms a cusp at $F=F^*$; different
    functions $ L(F)$ correspond to $P> P^*$ and $P< P^*$.

    Another kind of branching occurs at extrema of $\H(P)$, if any. There
    $F(P)$ behaves generically as $(P-P^*)^2$ while $ L_F\to \infty$, and a
    graph of $ L(F)$ smoothly touches the vertical axis $F=0$.
    The number of Lagrangians on the way from infinity to the center equals
    the number of monotonicity ranges of $F(P)$.

    All this is readily seen for specific examples. A qualitative picture
    for the choice (\ref{g2}) is shown in Fig.\,1.

    In the simplest case when $\H(P)$ is monotonic (e.g., like $\tanh P$),
    $ L(F)$ has only two branches $OP_1$ and $P_1 P_2$, and $P_2$ already
    corresponds to $r=0$.

    Thus any regular electric solution, being well-behaved with respect to
    the field equations in the $P$ framework, corresponds to different
    Lagrangians in different parts of space. This problem is absent for
    magnetic solutions since they are obtained directly in terms of $L(F)$.

\section{Comparison of effective metrics}

    The troubles with the electric solutions concern only the properties of
    NED, while the metric is quite well-behaved. The same is true for the
    electric field $F_{01}$. However, termination of a theory with given $
    L(F)$ implies some violent electromagnetic phenomena. For their
    understanding let us consider the effective metric introduced by Novello
    {\it et al. } \cite{nov1,nov2}
\beq                                                           \label{geff}
    h\MN= g\MN\eff = g\MN  L_F - 4 L_{FF} F^\mu{}_\alpha F^{\alpha\nu}.
\eeq
    As shown in \cite{nov1,nov2}, NED photons propagate along null
    geodesics of this metric.  For the space-time metric (\ref{M}), with a
    purely electric field the effective metric reads
\bear
        ds^2\eff\eql  h\mn dx^\mu dx^\nu =         	\label{dsefe}
    \frac{1}{\Phi}
              \biggl[ A(r) dt^2 - \frac{dr^2}{A(r)} \biggr]
                                        - \frac{r^2}{ L_F}d\Omega^2, \nn
        \Phi \eql  L_F + 2F  L_{FF} = \H_P/F_P.
\ear
    At an extremum $P=P^*$ of $F(P)$ where $F\ne 0$ (in particular, at
    the inevitable first minimum) one has $\Phi\to 0$ since $F_P\to 0$ while
    $\H_P$ is finite.  This leads to a curvature singularity of the
    effective metric, at least if $P^*$ is not located on
    a horizon, $A\ne 0$. Another kind of singularity of the metric
    (\ref{dsefe}) accompanies possible extrema of $\H(P)$. All
    this is verified by calculating the Kretschmann scalar $K$. Even more
    importantly, according to \cite{nov2}, if a NED photon comes from
    an emitter at rest at point 1 to an observer at rest at point 2, the
    corresponding frequencies $f_1$ and $f_2$ are related by
\beq
    \frac{f_2}{f_1} =
        \frac{h_{tt}(1)}{\sqrt{g_{tt}(1)}}\biggr/
        \frac{h_{tt}(2)}{\sqrt{g_{tt}(2)}}                  \label{z}
        =
        \frac{\Phi(2)/\sqrt{A(2)}}{\Phi(1)/\sqrt{A(1)}}
\eeq
    where the second equality corresponds to the metric (\ref{dsefe}).
    If $\Phi(2)=\infty$ [as it happens at a termination point of $L(F)$],
    then photons coming there are infinitely blueshifted and one may expect
    that they eventually lead to a real space-time singularity.

    For a magnetic solution, instead of (\ref{dsefe}), we get
\beq
        ds^2\eff =                                     \label{dsefm}
    \frac{1}{ L_F}
              \biggl[ A(r) dt^2 - \frac{dr^2}{A(r)} \biggr]
                                        - \frac{r^2}{\Phi}d\Omega^2
\eeq
    where again $\Phi  =  L_F + 2F L_{FF}$. Instead of (\ref{z}),
\beq
    \frac{f_2}{f_1}                                    	\label{zm}
        =
        \frac{ L_F(2)/\sqrt{A(2)}}{ L_F(1)/\sqrt{A(1)}}.
\eeq
    At the center ($r=0,\ A=1$) both $ L_F$ and $\Phi$ vanish,
    the coefficient $h_{22}\to \infty$, i.e., behaves as if in a
    wormhole, whereas $h_{00} \to \infty$, which means that photons
    arriving there, if any, would be infinitely redshifted [see \eq
    (\ref{zm})]. Actually, photons cannot reach a place where $ L_F=0$, as
    can be seen from an integral of their geodesic equation
\beq
     L_F^{-2} \dot r{}^2  + [A(r)\Phi/r^2] l^2 = \epsilon^2   \label{geo}
\eeq
    where the overdot is a derivative in the affine parameter, and
    $\epsilon$ and $l$ are the photon's constants of motion characterizing
    its initial energy and angular momentum. All curvature invariants
    of the metric (\ref{dsefm}) vanish at $r=0$. It is indeed a perfectly
    quiet place despite an infinite $F$.

    Some peculiarities, however, occur on the way from infinity to the
    center. There is always a sphere $r=r^*$ on which $\Phi=0$. It can be
    seen as follows: $\Phi$ may be represented as
    $\Phi = 2\sqrt{F} (\sqrt{F} L_F)_F$.  The quantity $\sqrt{F} L_F$
    vanishes at both $r=0$ and $r=\infty$ and is nonzero between them,
    hence it has at least one extremum at $F\ne 0$ --- this is where
    $\Phi=0$.  The metric (\ref{dsefm}), due to blowing-up of the coordinate
    spheres, has a singularity there, but the latter is actually unnoticed
    by NED photons, as is evident from \eq (\ref{geo}). Generically $ L_F\ne
    0$ where $\Phi=0$, therefore the photon frequency also remains finite.
    The meaning of the very fact of a curvature singularity of the effective
    metric is yet to be understood.

    If $ L_F=0$ at some $F>0$, this also causes a singularity of \eq
    (\ref{dsefm}), which acts as a potential wall (mirror) for NED photons,
    as is seen from (\ref{geo}). Accordingly, (\ref{zm})
    shows that they are infinitely redshifted: $f_2$ vanishes if
    $ L_F(2)=0$.  No photons from outside can thus approach the center.

    All this is in striking contrast to the picture obtained for an
    electric source --- we now have potential walls instead of wells and
    redshifts instead of blueshifts.

\section{Example}

    To have a specific example of a regular magnetic solution, let us
    employ the above $FP$ duality and consider, with slight modifications,
    the dependence (\ref{g2}), substituting $-\H$ with $L$ and $-P$ with $F$.
    An advantage of \eq (\ref{g2}) [as well as (\ref{g1}) and (\ref{g3})] is
    that it leads to a closed form of $M(r)$ and $A(r)$. Let us, however,
    slightly modify it, excluding an explicit dependence of $L$ on $m$ and
    $q$: they should be integration constants, while $ L$ may only contain
    fundamental constants or those originating from a deeper underlying
    theory. Moreover, to be able to describe systems with both electric and
    magnetic fields, where $F$ (and $P$) can have both signs, let us
    replace $-P$ with $|F|$ rather than $F$. So we put
\beq
     L(F) = F/\cosh^2 \bigl( a|F/2|^{1/4} \bigr), \quad a = \const.
                                                               \label{g2m}
\eeq
    The use of $|F|$ violates analyticity of $L$ at $F=0$:
    as required, $ L_F(0)=1$, but $ L_{FF}$ contains the discontinuous term
    $-a^2 |2F|^{-1/2}\sign F$. Though, in the range of interest, $F>0$,
    this $ L(F)$ is well-behaved. Integration in \eq (\ref{ML}) gives for a
    regular solution
\beq                                                           \label{Mg2}
    M(r) = \frac{|q|^{3/2}}{2a}
    		\biggl(1 - \tanh \frac{a\sqrt{|q|}}{r}\biggr),
\eeq
    so that $m = M(\infty) = |q|^{3/2}/2a$ ($q=\qm$), and some relations
    from \cite{gar2} are formally restored. In particular, the minimum
    value of $A(r) = 1- 2M(r)/r$ [recall that $A(0)=1$] depends on the
    ratio $\xi=m/|q|$, so that $A_{\min}$ is negative for $\xi > \xi_0
    \approx 0.96$ (we deal with a \bh\ with two horizons), zero for
    $\xi=\xi_0$ (an extremal \bh\ with one double horizon) and positive for
    $\xi < \xi_0$ (a regular particle-like system). It is of interest
    that, given any specific value of the constant $a$ in \eq (\ref{g2m}),
    we can obtain all three types of solutions depending on the charge
    value:  we have a non-extremal or extremal \bh\ if $|q| \leq
    4a^2/\xi_0^2$, or we have a particlelike solution (a monopole)
    otherwise.  Despite the restriction imposed by the regularity condition,
    one finds all three types of regular solutions. This feature seems to be
    quite generic for proper nonlinear Lagrangians. One can also verify that
    the properties of the effective metric (\ref{dsefm}) confirm the above
    general observations.

    One can also notice that, due to an exponential decay of $M(r)$ in
    \eq (\ref{Mg2}), the metric is in this case $C^\infty$ smooth at $r=0$.

\section{Concluding remarks}

    A more complete description of the properties of the present regular NED
    solutions, as well as others, requires a better understanding of the
    long-standing and non-trivial problem of motion of charged bodies in
    NED, probably following the lines of Refs. \cite{force,ryb,dirac51}.

    One more subject of interest for further study is the inclusion of
    another electromagnetic field invariant, ${}^*F\MN F\mn$, into the
    Lagrangian in addition to $F$. This invariant is involved, in
    particular, in the Born-Infeld and Heisenberg-Euler NED Lagrangians; its
    appearance should be able to widen the diversity of regular \bh\ and
    monopole solutions. Related subjects are the $FP$ duality
    between solutions of different theories involving both invariants and a
    possible extension of this duality to non-\sph\ configurations.

\medskip\noi
{\bf Acknowledgments.}
    I acknowledge the partial financial support of the Russian Ministry of
    Science and Technologies and DFG Proj. 436 RUS 113/236/0(R) and the kind
    hospitality at Potsdam University.  I am sincerely grateful to Irina
    Dymnikova, Yuri Rybakov, Vitaly Melnikov, Vladimir Dzhunushaliev, Gernot
    Neugebauer, Claus L\"ammerzahl, Hans-J\"urgen Schmidt and Martin Rainer
    for helpful discussions.

\end{document}